\documentclass{aastex}
\usepackage{spr-astr-addons} 
\usepackage{multicol}
\usepackage{amsmath}
\usepackage{amsfonts}
\usepackage{amssymb}
\usepackage{graphicx}
\usepackage{hyperref}
\hypersetup{
colorlinks = true, allcolors=blue}
\usepackage{colordvi}
\usepackage{latexsym}
\usepackage{url}
\usepackage{lipsum}
\usepackage{mathtools}
\usepackage{cuted}

\newcommand{\emaila}{bibekanandafm@email.com}

\begin{document}
	
\title{Generalized Interacting Dark Energy Model and Loop Quantum Cosmology}
\author{Suryakanta Swain\altaffilmark{1}} 
\author{Debasis Sahu\altaffilmark{1}}
\author{Debabrata Dwivedee\altaffilmark{2}}
\author{Gourishankar Sahoo\altaffilmark{3}}
\author{Bibekananda Nayak\altaffilmark{1*}}
\email{\emaila} 
\altaffiltext{1}{P.G. Department of Applied Physics and Ballistics, Fakir Mohan University, Balasore, Odisha 756019, India}
\altaffiltext{2}{Department of Physics, Prananath College (Autonomous), Khordha, Odisha 752057, India}
\altaffiltext{3}{Department of Physics and Astronomical Sciences, Central University of Himachal Pradesh, Dharamshala, 176206, India}
\altaffiltext{*}{E-mail: bibekanandafm@gmail.com}

\begin{abstract}
	The recently observed accelerated expansion of the universe has put a challenge for its theoretical understanding. As a possible explanation of this, it is considered that the most part of the present universe is filled with a form of energy that exerts a negative pressure called dark energy, which drives the acceleration. In the present work, we assume a dynamical dark energy model, where dark energy interacts with matter and grows at the expense of the latter. Using this model, we discuss the evolution of the universe within the context of loop quantum cosmology. Our work successfully explains the presently observed accelerated expansion of the universe, by predicting that the present universe is phantom dominated. We also found that in the past, the expansion of the universe was decelerated one and transition from deceleration to acceleration would occur at $t_{q=0}=0.688t_0$, where $t_0$ is the present age of the universe. Again, our analysis predicted that at the transition time, the universe would be dominated with quintessence type dark energy. 
\end{abstract}


\section{Introduction}
In theoretical physics, classical cosmology, which is described by the Einstein's general theory of relativity, solves many mysteries about the universe. General theory of relativity provides the idea about gravitational force and its implications, in cosmological and astronomical scales. But, it is unable to give a reasonable explanation on Big-Bang singularity in macroscopic scale. So we would like to orient our focus from classical aspect to the quantum one, which deals with the physics of matter in small scales. This general theory of relativity and quantum theory, are the two excellent proven theories of modern physics. However, both these individual theories cannot successfully explain the complications like Big-Bang singularity that arise in standard cosmology. A solution to that may be achieved by combining both the theories into a single theory, coined as quantum gravity theory. Loop Quantum Gravity (LQG) \citep{Ash2004, Rovelli2004, Thiemann2007, Han2007, Bod2016, Chi2015, Ash2012} is one of the special alluring features of quantum gravity theories. LQG is completely non-perturbative, explicit background independent approach to quantum gravity. Generally, implication of LQG on cosmology for the survey of our universe, is called loop quantum cosmology (LQC) \citep{Ash2006a, Ash2006b, Ash2011, Banerjee2012}. The several outcome of LQC can be explained on the basis of two possible methods. The first kind of methods introduce modifications in the inverse scale factor below the critical scale factor. These methods also provide the idea about the replacement of classical Big-Bang by quantum bounce with some advantages of avoidance of singularities \citep{Sam2006}, inflation related problems \citep{Zha2007, Bojowald2011} etc. But when the scale factor is very large, the modifications introduced by first kind of methods become ineffective, and then the second kind of methods come into picture \citep{Che2008, Jam2011}. The second kind of methods add a quantum correction term to the standard Friedmann equation \citep{Ash2006a, Ash2006b, Singh2006, Copeland2006, Art2009}. This term is $-\frac{\rho^2}{\rho_c}$, where $\rho$ is the density of mass-energy of the universe and $\rho_c$ is the critical density at which bounce occurs. In a contracting universe, $\dot{a}$ is less than zero. But when the density of mass-energy ($\rho$) is the same order as critical density $\rho_c$,  $\dot{a}$ becomes zero, which indicates the end of contraction of the universe \citep{Bag2018}. Hence the universe moves from the classical Big-Bang to the quantum big bounce and oscillates forever. Again, this kind of modification shows the avoidance of Big-Bang singularity, Big Rip \citep{Alo2018, Har2012} and several future singularities \citep{Sam2006}. In LQC, several features like quantum bounce \citep{Zhu2017}, super inflation \citep{Xi2013} and future singularity problem etc., attract cosmologists and mathematicians to a common platform to make out our own universe. Also, recently more researchers have paid attention for explaining presently observed accelerated expansion of the universe in LQC \citep{Cognola2005, Fu2008, Wu2008, Xiao2010, Jami2010, Sadjadi2011, Oik2019}.

On the other hand, observations \citep{Ade2014, Ade2016a, Ade2016b} demand that present universe is largely dominated by two components with unknown character and origin, named as dark energy and dark matter. In some models, it is assumed that perhaps the dark energy and dark matter are coupled to each other, so that they behave like a single dark fluid. Although this consideration sounds slightly phenomenological, but this possibility cannot be ruled out by any observations. So, one can of course think of some interaction between these two fields. The idea of coupling in the dark sectors was initiated by Wetterich \citep{Wetterich1995} and subsequently discussed by Amendola \citep{Amendola2000} and others. In the last couple of years, different kind of interacting dark energy models \citep{Khu2017, Xu2016, Zim2012, Maj2010, Yan2017, Yan2018a, Nay2019, Yan2018b, Bam2012, Mis2019a, Ray2019, Mis2019b, Nay2009, Bil2000, Oli2005, del2009, Val2010, Pan2013, Yan2014, Pan2015, Muk2017, Sah2019, Odi2017, Odi2018} have been studied by several researchers. But most of the studies on the dark energy models, are conducted in the light of classical Einstein gravity. However, the evolution of the universe seems plausible, if one considers quantum gravity. So the study on impact of interacting dark energy in the framework of quantum gravity will become a useful tool to understand the dynamical universe.

In this work, we focus on the evolution of the universe within the context of loop quantum cosmology by assuming a non-gravitational interaction between dark energy and dark matter. Here, we aim to discuss the variations of cosmic scale factor, matter-energy density and deceleration parameter for different cosmological eras on the basis of LQC. In our study, we also try to find out the exact form of dark energy for the present universe. Again in this environment, we investigate the form of dark energy at transition period, where the expansion of the universe goes from decelerated to accelerated one. Comparing these two stages of evolution, we try to describe the nature of dark energy through its equation of state parameter, for whole evolution of the universe.

\section{Basic Framework}

For a spatially flat FRW universe ($k=0$) filled with dust and dark energy, the Friedmann equation, Raychaudhuri's equation and energy conservation equation in LQC \citep{Bojowald2005, Ashtekar2006c, Che2008, Jam2011, Li2010} take the form 
\begin{equation} \label{friedmann}
	H^2=\frac{8\pi G}{3}\left(\rho_x+\rho_m\right)\left\{1-\frac{(\rho_x+\rho_m)}{\rho_c}\right\},
\end{equation}
\begin{equation} \label{raychaudhuri}
	\dot{H}=-4\pi G\left(\rho_x+\rho_m+p_x\right)\left\{1-\frac{2(\rho_x+\rho_m)}{\rho_c}\right\},
\end{equation}
and
\begin{equation} \label{energy cons.}
	\left(\dot\rho_x+\dot\rho_m\right)+3H\left(\rho_x+\rho_m+p_x\right)=0
\end{equation}
respectively.
Where $H=\frac{\dot{a}}{a}$ is the Hubble parameter, $\rho_x=$ dark energy density, $\rho_m=$ matter density, $p_x=$ pressure of the dark energy and $\rho_c$ represents the critical value of energy density of the universe given by $\rho_c=\frac{\sqrt{3}}{16\pi^2 \gamma^3} \rho_{pl}$ with $\gamma=\frac{\ln2}{\pi\sqrt{3}}$ is the dimensionless Barbero-Immirzi parameter \citep{Ash1998, Dom2004, Mei2004} and $\rho_{pl}$ is the energy density of the universe in Plank time.

In generalized interacting dark energy model, we assume that during evolution of the universe, dark energy and matter do not conserve separately but they interact with each other and one may grow at the expense of the other. Inserting this idea, energy conservation equation can be written as 
\begin{eqnarray} \label{energy-split}
	\dot\rho_m+3H\rho_m=Q, \nonumber \\
	\dot\rho_x+3H\left(1+\omega\right)\rho_x=-Q,
\end{eqnarray}
where $Q=\Gamma\rho_x$ with $\Gamma>0$, is the interaction rate \citep{Sen2008}, having dimension of Hubble parameter and $\omega=\frac{p_x}{\rho_x}$ denotes the equation of state parameter for dark energy. But the observations \citep{Ade2016a} demand that $68.3\%$ of present universe is filled with dark energy and present age of the universe is $13.82{\times{10^9}}$ years. Again, by considering that the early universe is completely filled with dust, and dark energy appeared due to its decay with $\Gamma$ as the interaction rate, we can estimate $\Gamma$ as \citep{Nay2019}
\begin{equation} \label{Gamma value}
	\Gamma \approx 4.942{\times{10^{-11}}} (yr)^{-1}.
\end{equation}
Thus at any time $t$, the ratio of matter to dark energy density becomes
\begin{equation} \label{r-expression}
	r=\frac{\rho_m}{\rho_x}=\frac{1-\Gamma t}{\Gamma t}.
\end{equation}

\section{Deceleration Parameter}

The cosmological parameter, which determines the nature of expansion of the universe, is known as deceleration parameter and mathematically, it can be written as $q=-1-\frac{\dot{H}}{H^2}$. But from equations (\ref{friedmann}) and (\ref{raychaudhuri}), we have 
\begin{equation} \label{ratio}
	\frac{\dot{H}}{H^2}=-\frac{3}{2}\left(\frac{1+r+\omega}{1+r}\right)\left[\frac{1-2\left(\frac{\rho_m+\rho_x}{\rho_c}\right)}{1-\left(\frac{\rho_m+\rho_x}{\rho_c}\right)}\right].
\end{equation}
So deceleration parameter $q$ becomes
\begin{equation} \label{q1}
	q=-1+\frac{3}{2}\left(\frac{1+r+\omega}{1+r}\right)\left[\frac{1-2\left(\frac{\rho_m+\rho_x}{\rho_c}\right)}{1-\left(\frac{\rho_m+\rho_x}{\rho_c}\right)}\right].
\end{equation}

According to our model at any time $t$, the density of the universe is  $\rho=\rho_m+\rho_x$ and in LQC, the expression for $\rho$ in radiation-dominated and matter-dominated eras can be  written as \citep{Dwi2014}

\begin{align} \label{rho-r}
		\rho\left(t\right)_{t<t_e}=&\rho_0\Biggl[\frac{\rho_0}{\rho_c}+\Biggl\{2 \sqrt{\frac{8\pi G}{3}}\rho_0^{\frac{1}{2}}\left(t-t_e\right) \notag\\
		&+\frac{3}{2}\sqrt{\frac{8\pi G}{3}}\rho_0^{\frac{1}{2}}\left(t_e-t_0\right) \notag\\
		&+\left(1-\frac{\rho_0}{\rho_c}\right)^{\frac{1}{2}}\Biggr\}^2\Biggr]^{-1},
\end{align}
and 
\begin{align}\label{rho-m} 
		\rho\left(t\right)_{t>t_e}=&\rho_0\Biggl[\frac{\rho_0}{\rho_c}+\Biggl\{\frac{3}{2}\sqrt{\frac{8\pi G}{3}}\rho_0^{\frac{1}{2}}\left(t-t_0\right) \notag\\
		&+\left(1-\frac{\rho_0}{\rho_c}\right)^{\frac{1}{2}}\Biggr\}^2\Biggr]^{-1},
\end{align}
where $t_e$ is the time of radiation-matter equality and $t_0$ is the present age of the universe.

Now deceleration parameter can be written as
\begin{equation} \label{q2}
	q=-1+\frac{3}{2}\left(\frac{1+r+\omega}{1+r}\right)\left[\frac{1-2\left(\frac{\rho}{\rho_c}\right)}{1-\left(\frac{\rho}{\rho_c}\right)}\right].
\end{equation}
At $\rho=\frac{\rho_c}{2}$, we get $q=-1$, which indicates the occurrence of inflation in the early universe \citep{Guth1981, Kofman1994}. Because during inflation $a\propto e ^ {\alpha t}$ and hence $q=-1$. Again for accelerated expansion, the second term of R.H.S. in equation (\ref{q2}) should be less than $1$ i.e. 
\begin{equation}
	\frac{3}{2} \left(\frac{1+r+\omega}{1+r}\right)\left[\frac{1-\frac{2\rho}{\rho_c}}{1-\frac{\rho}{\rho_c}}\right]<1 , \nonumber	
\end{equation}
which gives
\begin{equation} \label{omega}
	\omega<-\left(1+r\right)+\frac{2}{3}\left(1+r\right)\left[\frac{1-\frac{\rho}{\rho_c}}{1-\frac{2\rho}{\rho_c}}\right].
\end{equation}
Thus $\omega$ is negative and its magnitude must be
\begin{equation}
	\lvert\omega \rvert>\left(1+r\right)-\frac{2}{3}\left(1+r\right)\left[\frac{1-\frac{\rho}{\rho_c}}{1-\frac{2\rho}{\rho_c}}\right]\nonumber.
\end{equation}
For present accelerated expansion
\begin{equation} \label{omega mod}
	\lvert\omega \rvert>\left(1+r_0\right)-\frac{2}{3}\left(1+r_0\right)\left[\frac{1-\frac{\rho_0}{\rho_c}}{1-\frac{2\rho_0}{\rho_c}}\right].
\end{equation}
Using the values of $r_0$, $\rho_0$ and $\rho_c$ in above equation (\ref{omega mod}), we get 
\begin{equation} \label{omega mod value}
	\lvert\omega \rvert>0.488.
\end{equation}

In our calculation, we have taken $\rho_0=1.1{\times{10^{-29}}}$ $(gm/cm^{3})$,  $\rho_c=5.317{\times{10^{94}}}$ $(gm/cm^{3})$, $G=6.673$ ${\times{10^{-8}}}$ $(dyne-cm^{2}/gm^{2})$, $t_0=4.36{\times{10^{17}}}$  $(sec)$, $\Gamma \approx 4.942$ ${\times{10^{-11}}}$ $(yr)^{-1}$.

Now we construct Table \ref{table1} from equation (\ref{q2}), which shows the variation of deceleration parameter ($q$) with equation of state parameter ($\omega$) of dark energy, for present time ($t_0$).

\begin{table}[ht]
	\caption{Variation of deceleration parameter ($q$) with equation of state parameter of dark energy ($\omega$), for present time ($t_0$)}
	\label{table1}
	\begin{tabular}{|c|c||c|c|}
		\hline	
		$\omega$ & $q$ & $\omega$ & $q$\\
		\hline
		-0.50 & -0.0124 &  -1.02449 & -0.5500\\
		\hline
		-0.55 & -0.0636 &  -1.05 & -0.5761\\
		\hline
		-0.60 & -0.1149 &  -1.10 & -0.6273\\
		\hline
		-0.65 & -0.1661 & -1.15 & -0.6786\\
		\hline
		-0.70 & -0.2174 & -1.20 & -0.7298\\
		\hline
		-0.75 & -0.2686 & -1.25 & -0.7811\\
		\hline
		-0.80 & -0.3199 & -1.30 & -0.8323\\
		\hline
		-0.85 & -0.3711 & -1.35 & -0.8836\\
		\hline
		-0.90 & -0.4224 &  -1.40 & -0.9348\\
		\hline
		-0.95 & -0.4736 & -1.45 & -0.9861\\
		\hline
		-1.00 & -0.5249 &  -1.50 & -1.0373\\
		\hline
	\end{tabular}
\end{table}

Again, by comparing with present observational data \citep{San2016}, $q_0\approx-0.55$, we can get $\omega_0=-1.02449$. Here, subscript `$0$' refers to present value.

For the radiation-dominated era($t<t_e$), the scale factor varies as \citep{Dwi2014} 
\begin{align} \label{a-r}
	a\left(t\right)_{t<t_e}=&\Biggl[\frac{\rho_0 a_0^3 a_e}{\rho_c}+\Biggl\{2\rho_0^\frac{1}{2}a_0^\frac{3}{2}a_e^\frac{1}{2}\sqrt{\frac{8\pi G}{3}}\left(t-t_e\right) \notag\\
	&+\left(a_e^4-\frac{\rho_0 a_0^3 a_e}{\rho_c}\right)^\frac{1}{2}\Biggr\}^2\Biggr]^\frac{1}{4}.
\end{align}
Similarly for the matter-dominated era($t>t_e$), the scale factor varies as
\begin{align} \label{a-m}
	a\left(t\right)_{t>t_e}=&\Biggl[\frac{\rho_0 a_0^3}{\rho_c}+\Biggl\{\frac{3}{2}\rho_0^\frac{1}{2}a_0^\frac{3}{2}\sqrt{\frac{8\pi G}{3}}\left(t-t_0\right) \nonumber\\
	&+\left(a_0^3-\frac{\rho_0 a_0^3}{\rho_c}\right)^\frac{1}{2}\Biggr\}^2\Biggr]^\frac{1}{3}.
\end{align}
Thus, for matter-dominated era, redshift ($z$) can be calculated as 
\begin{align} \label{redshift}
	z=&\Biggl[\frac{\rho_0}{\rho_c}+\Biggl\{\frac{3}{2}\rho_0^\frac{1}{2}\sqrt{\frac{8\pi G}{3}}t_0\left(\frac{t}{t_0}-1\right) \nonumber\\
	&+\left(1-\frac{\rho_0}{\rho_c}\right)^\frac{1}{2}\Biggr\}^2\Biggr]^{-\frac{1}{3}}-1.
\end{align}
So the expression for $(t/t_0)$ can be written in terms of $z$ as
\begin{equation} \label{redshift2}
	\frac{t}{t_0}=1+
	\left[\frac{[(1+z)^{-3}-\frac{\rho_0}{\rho_c}]^{\frac{1}{2}}-[1-\frac{\rho_0}{\rho_c}]^{\frac{1}{2}}}{{\frac{3}{2}}\rho_0^{\frac{1}{2}}\sqrt{\frac{8\pi G}{3}}t_0}\right].
\end{equation}
	
If we consider the present universe is dark energy dominated, then the dark energy domination would start from the time, at which transition from deceleration to acceleration would occur. At transition time, $q=0$ and thus equation (\ref{q2}) by the use of equation (\ref{r-expression}), gives
\begin{equation} \label{omega q=0}
	\omega_{q=0}=\frac{2}{3\Gamma t}\left(\frac{1-\frac{\rho}{\rho_c}}{1-\frac{2\rho}{\rho_c}}\right)-\frac{1}{\Gamma t}.
\end{equation}
Using equations (\ref{redshift}) and (\ref{omega q=0}) we construct Table \ref{table2}, which shows variations of transition redshift ($z_{q=0}$) and transition equation of state parameter of dark energy ($\omega_{q=0}$) with time.

\begin{table}[ht]
	\caption{Variation of transition redshift ($z_{q=0}$) and transition equation of state parameter of dark energy ($\omega_{q=0}$) with time}
	\label{table2}
	\begin{tabular}{|c|c|c|}
		\hline	
		$\frac{t}{t_0}$ & $z_{q=0}$ & $\omega_{q=0}$\\
		\hline
		1.0 & 0 & -0.4878\\
		\hline
		0.95 & 0.0580 & -0.5135\\
		\hline
		0.90 & 0.1252 & -0.5420\\
		\hline
		0.85 & 0.2042 & -0.5739\\
		\hline
		0.80 & 0.2987 & -0.6098\\
		\hline
		0.75 & 0.4143 & -0.6504\\
		\hline
		0.70 & 0.5596 & -0.6969\\
		\hline
		0.6880 & 0.6003 & -0.70908\\
		\hline
		0.65 & 0.7489 & -0.7505\\
		\hline
		0.60 & 1.0087 & -0.8130\\
		\hline
		0.55 & 1.3929 & -0.8870\\
		\hline
		0.50 & 2.0359 & -0.9757\\
		\hline
	\end{tabular}
\end{table}

In comparison with the present observation \citep{San2016} $z_{q=0}\approx 0.6$, one can get from our analysis that $t_{q=0}\approx 0.688t_0$ and $\omega_{q=0}=-0.70908$.

\section {Nature of Dark Energy and Accelerated Expansion}
As we have seen from previous section, equation of state parameter of dark energy $\omega$ varies with time in such a way that at $t=0.688t_0$, $\omega=-0.70908$ and at $t=t_0$, $\omega=-1.02449$. Now for constructing the exact form of equation of state parameter of dark energy, we consider $\omega$ in a general quadratic form as
\begin{equation}
	\omega=at^2+bt\nonumber
\end{equation}
where, $a$ and $b$ are the constant coefficients. Here, we have taken the quadratic form of $\omega$. Because we have only two values of $\omega$ for two different times and so we have the possibilities to take $\omega$ as either a straight line or a quadratic function. But quadratic function contains more variation than a straight line and hence it is more appropriate for an unknown function, to be chosen as a quadratic function.

For the present era, $\omega$ takes the form
\begin{equation} \label{omega eq.1}
	-1.02449=at_0^2+bt_0.
\end{equation}
And similarly for the recent past, when the deceleration parameter is zero, $\omega$ takes the form 
\begin{equation} \label{omega eq.2}
	-0.70908=a\left(0.688\right)^2t_0^2+b\left(0.688\right)t_0.
\end{equation}
Solving Equations (\ref{omega eq.1}) and (\ref{omega eq.2}), we get
\begin{center}
	$a=\frac{0.01971}{t_0^2}$ and $	b=-\frac{1.0442}{t_0}$.
\end{center}
Putting the values of $a$ and $b$ in the expression of $\omega$, we get 
\begin{equation} \label{omega expression}
	\omega=0.01971\left(\frac{t}{t_0}\right)^2-1.0442\left(\frac{t}{t_0}\right).
\end{equation}

\begin{figure}[tb]
	\includegraphics[scale=0.65]{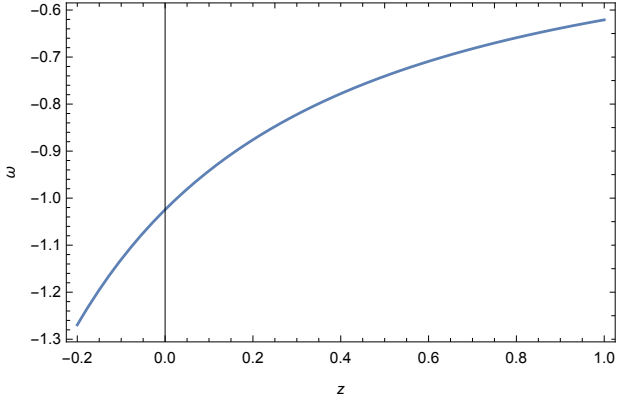}
	\caption{Variation of equation of state parameter of dark energy ($\omega$) with redshift($z$)}
	\label{fig1}
\end{figure}

We plot Figure \ref{fig1} from equation (\ref{omega expression}) by using equation (\ref{redshift2}), which shows the variation of equation of state parameter of dark energy ($\omega$) with redshift ($z$).
From Figure \ref{fig1}, we found that $\omega$ $<$ $-1$ for present time ($z=0$), which indicates that the present universe is phantom dominated.

\begin{figure}[tb]
	\includegraphics[scale=0.65]{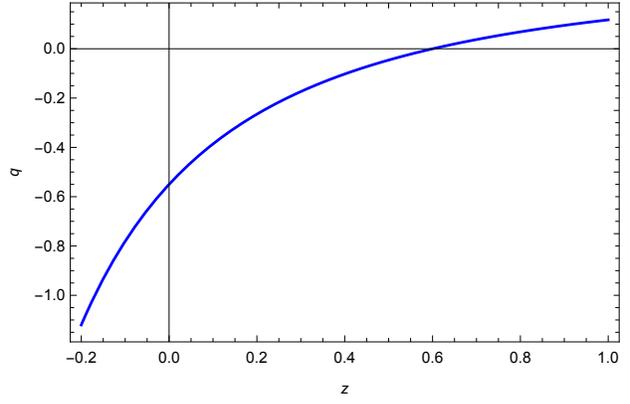}
	\caption{Variation of deceleration parameter ($q$) with redshift ($z$)}
	\label{fig2}
\end{figure}

The equation for deceleration parameter can be written by using equations (\ref{r-expression}), (\ref{rho-m}) and (\ref{omega expression}) in equation (\ref{q2}), as
\begin{align} \label{q expression}
	&q=-1+\frac{3}{2} \notag\\
	&\left\{\frac{1+\left(\frac{1-\Gamma\left(\frac{t}{t_0}\right)t_0}{\Gamma\left(\frac{t}{t_0}\right)t_0}\right)+0.01971\left(\frac{t}{t_0}\right)^2-1.0442\left(\frac{t}{t_0}\right)}{1+\left(\frac{1-\Gamma\left(\frac{t}{t_0}\right)t_0}{\Gamma\left(\frac{t}{t_0}\right)t_0}\right)}\right\} \notag\\
	&\left[\frac{1-2\left[\frac{\rho_0\left[\frac{\rho_0}{\rho_c}+\left\{\frac{3}{2}\sqrt{\frac{8\pi G}{3}}\rho_0^{\frac{1}{2}}\left(t-t_0\right)+\left(1-\frac{\rho_0}{\rho_c}\right)^{\frac{1}{2}}\right\}^2\right]^{-1}}{\rho_c}\right]}{1-\left[\frac{\rho_0\left[\frac{\rho_0}{\rho_c}+\left\{\frac{3}{2}\sqrt{\frac{8\pi G}{3}}\rho_0^{\frac{1}{2}}\left(t-t_0\right)+\left(1-\frac{\rho_0}{\rho_c}\right)^{\frac{1}{2}}\right\}^2\right]^{-1}}{\rho_c}\right]}\right].
\end{align}
We plot Figure \ref{fig2} from equation (\ref{q expression}) by using equation (\ref{redshift2}), which shows the variation of deceleration parameter ($q$) with redshift ($z$).
From Figure \ref{fig2}, we found that the present universe undergoes an accelerated phase of expansion and the transition from decelerated to accelerated expansion of the universe would be occurred at $z_{q=0} \approx 0.6$ (or $t_{q=0}\approx 0.688t_0$).  

\section{Discussion and Conclusion}
We, here, have studied the evolution of the universe in LQC by using a dynamical interacting dark energy model, where dark energy interacts with matter and grows at the expense of latter. First we used the concept of interacting dark energy in evaluating the deceleration parameter for different cosmic eras. From our analysis, we found that for explaining the accelerated expansion, the present universe must be dominated by phantom type dark energy, since the value of equation of state parameter of dark energy is found to be $-1.0245$. Again we have picturised the nature of dark energy through evaluating its equation of state parameter for whole evolution of the universe. To get this, we first calculated the transition time and found that the transition from deceleration to acceleration would be occurred, when the age of the universe was nearly $0.688$ times the present age. At that time, the value of equation of state parameter of dark energy would be $-0.709$, which shows that the universe was then dominated with quintessence type of dark energy. Comparing these two stages of evolution, one at present time and another at transition time, we construct an expression for equation of state parameter of dark energy for whole evolution of the universe. Using this, we have shown the variation of deceleration parameter with redshift for whole evolution of the universe and from our analysis, we concluded that the universe will undergo an accelerated phase of expansion even in far future.

\section*{Data Availability Statement}
No new data were created or analyzed in this study. Data sharing is not applicable to this article.

\section*{Ethics Declarations}
\subsection*{Conflict of Interest}
The authors declare that they have no conflicts of interest.
%

\end{document}